\documentstyle[twoside,12pt,epsfig]{article}
\begin{document}

\newcommand{\muA}{\mbox{\,$\mu\!$A\ }}
\newcommand{\GeVx}{\mbox{\,Ge$\!$V}}
\newcommand{\GeV}{\mbox{\,Ge$\!$V\ }}
\newcommand{\MeVx}{\mbox{\,Me$\!$V}}
\newcommand{\MeV}{\mbox{\,Me$\!$V\ }}
\newcommand{\mumx}{\mbox{\,$\mu$m}}
\newcommand{\mum}{\mbox{\,$\mu$m\ }}
\newcommand{\mmx}{\mbox{\,mm}}
\newcommand{\mm}{\mbox{\,mm\ }}
\newcommand{\nbarn}{\mbox{\,nb\ }}
\newcommand{\nbarnx}{\mbox{\,nb}}
\newcommand{\mubarn}{\mbox{\,$\mu$b\ }}
\newcommand{\mubarnx}{\mbox{\,$\mu$b}}
\newcommand{\dsTx}{\mbox{\,$\Delta\sigma_{\rm \small T}$}}
\newcommand{\dsT}{\mbox{\,$\Delta\sigma_{\rm \small T}$\ }}
\newcommand{\dsLx}{\mbox{\,$\Delta\sigma_{\rm \small L}$}}
\newcommand{\dsL}{\mbox{\,$\Delta\sigma_{\rm \small L}$\ }}
\newcommand{\sigtotx}{\mbox{\,$\sigma_{\rm tot}$}}
\newcommand{\sigtot}{\mbox{\,$\sigma_{\rm tot}$\ }}
\newcommand{\perc}{\mbox{\,\%\ }}
\newcommand{\percx}{\mbox{\,\%}}
\newcommand{\grad}{\mbox{$^{\circ}$}\ }
\newcommand{\gradx}{\mbox{$^{\circ}$}}
\newcommand{\cmx}{\mbox{\,cm}}
\newcommand{\cm}{\mbox{\,cm\ }}
\newcommand{\sx}{\mbox{\,s}}
\newcommand{\s}{\mbox{\,s\ }}

\title{\LARGE Measurement of Partial-Wave Contributions\\
       in $pp \rightarrow pp\pi^0$}

\author{ 
H.O.\ Meyer, J.T.\ Balewski, J.\ Doskow, R.E.\ Pollock, B.\ v.\ Przewoski \\
T.\ Rinckel, P.\ Th\"orngren--Engblom, A.\ Wellinghausen \\
{\small \it Dept.\ of.\ Physics and Cyclotron Facility, 
 Indiana University, Bloomington, IN 47405} \\[1.2ex]
W.\ Haeberli, B.\ Lorentz, F.\ Rathmann\footnote{Present address:  
    Forschungs Zentrum J\"ulich GmbH, 52425 J\"ulich, Germany.}, 
B.\ Schwartz and T.\ Wise \\
{\small \it University of Wisconsin-Madison, Madison, WI 53706} \\[1.2ex]
W.W.\ Daehnick and Swapan K.\ Saha\footnote{Permanent address: 
 Bose Institute, Calcutta 700009, India}\\
{\small \it Dept.\ of Physics and Astronomy, Univ.\ of Pittsburgh,
  Pittsburgh, PA 15260} \\[1.2ex]
P.V.\ Pancella \\
{\small \it Western Michigan University, Kalamazoo, MI 49008} } 

\maketitle

\begin{abstract}
We report a measurement of the spin-dependent total cross section ratios
\dsTx/\sigtot and \dsLx/\sigtot of the $pp \rightarrow pp\pi^0$
reaction between 325\MeV and 400\MeVx. The experiment was carried out 
with a polarized internal target in a storage ring. Non-vertical beam 
polarization was obtained by the use of solenoidal spin rotators. 
Near threshold, the knowledge of both spin--dependent total cross 
sections is sufficient to deduce the strength of certain participating
partial waves, free of any model.
\end{abstract}

\centerline{PACS numbers: 24.70.+s , 25.10.+s , 29.20.Dh , 20.25.Pj , 29.27.Hj}

\vspace{2mm}

We present a measurement of \dsTx/\sigtot and \dsLx/\sigtot for the 
$pp \rightarrow pp\pi^0$ reaction. The quantity \dsT (or \dsLx) 
equals the difference between the total cross sections measured with 
opposite and parallel, transverse (or longitudinal) beam and target 
polarizations, while \sigtot is the (unpolarized) total cross section. 
At threshold a single partial wave ($^3P_0 \rightarrow {^1S_0,l=0}$)
dominates. As the bombarding energy is increased other partial waves
become significant. In the following, we will demonstrate how the 
results of the present experiment can be used to gain information on 
these higher partial waves, making it possible to test theoretical 
models selectively.

The reaction amplitude for pion production in the nucleon-nucleon (NN)
system may be separated into contributions with definite angular
momentum. These partial-wave amplitudes are labelled with the quantum
numbers of the final state $(^{2{\rm S}+1}$L$_{\rm J},l)_{\rm j}$, 
where S, L and J are the spin, angular momentum and total angular
momentum of the NN
pair, $l$ is the angular momentum of the pion and j is the total 
angular momentum of the final state. The initial-state quantum 
numbers $\cal S$, $\cal L$, $\cal J$ are related to those of the 
final state by angular momentum, parity and isospin conservation. 

It is useful to define cross sections $^{2{\cal S}+1}\sigma_m$ which
partition the total cross section \sigtot according to the initial 
NN spin, $\cal S$, and its projection, $m$, onto the direction of the 
incident momentum. The three possible contributions $^1\sigma_0$, 
$^3\sigma_0$, and $^3\sigma_1$ are related to the observable 
quantities by
\begin{equation} \begin{array}{ll}
^1\sigma_0\ =\ \sigtot\ +\ \dsT\!\! & +\ \frac{1}{2}\dsL \\[1.5ex] 
^3\sigma_0\ =\ \sigtot\ -\ \dsT\!\! & +\ \frac{1}{2}\dsL \\[1.5ex]
^3\sigma_1\ =\ \sigtot\             & -\ \frac{1}{2}\dsL\ .
\end{array}
\end{equation}

Because, near threshold, the relative kinetic energies between 
the particles in the final state are small, the angular momenta 
L and $l$ are either 0 or 1, giving rise to the possible final 
states L$l$  = Ss, Sp, Ps, or Pp. A list of possible 
$pp \rightarrow pp\pi^0$ amplitudes can easily be constructed, 
taking into account the negative parity of the pion, as well as 
conservation of angular momentum and isospin. From this list, 
one sees that the Sp final state is not allowed, that only 
initial--state singlets ($^1\sigma_0$) contribute to Ps final states,
that ($^3\sigma_1$) contributes only to Pp, and that ($^3\sigma_0$) 
may contribute to both, Ss and Pp. Assuming that there are no final 
states other than Ss, Ps, and Pp, we require that \sigtot = 
$\sigma_{\rm Ss} + \sigma_{\rm Ps} + \sigma_{\rm Pp}$ where 
$\sigma_{{\rm L}l}$ is the total cross section to a certain final state,  
L$l$. We can then summarize the above in the following way:
\begin{equation} \begin{array}{l}
^1\sigma_0\ =\ 4 \sigma_{Ps} \\[1.2ex]
^3\sigma_0\ =\ 4 (\sigma_{Ss} + \hat{\sigma}_{Pp}) \\[1.2ex]
^3\sigma_1\ =\ 2 (\sigma_{Pp} - \hat{\sigma}_{Pp})\ .
\end{array}
\end{equation}

Here, $\sigma_{Pp}$ and $\hat{\sigma}_{Pp}$ represent two different 
combinations of Pp amplitudes. The term $\hat{\sigma}_{Pp}$
is needed in $^3\sigma_0$ because 
${\cal S}=1, m=0$ initial states can lead to Ss {\em and} Pp 
final states. Since the term $\hat{\sigma}_{Pp}$ does not contribute 
to the total cross section, it must also appear in $^3\sigma_1$. 
Combining Eqs.\ 1 and 2, we arrive at the useful relation 
\begin{equation}
 \frac{\sigma_{Ps}}{\sigtot}\ =\ \frac{1}{4}\left( 1 +
 \frac{\dsT}{\sigtot} + \frac{1}{2}\frac{\dsL}{\sigtot}\right)\ ,
\end{equation}
which states that the probability to form a Ps final state can be deduced,
free of any model, from the observables of this experiment. A similar
equation for $(\sigma_{Pp} - \hat{\sigma}_{Pp})/\sigtot$ follows easily
from Eqs.\ 1,2. 

The experiment discussed here was carried out with the Indiana Cooler.
Protons from the cyclotron were stack-injected into the ring at 197\MeVx,
reaching an orbiting current of several 100\muA within a few minutes. 
The beam was then accelerated to the energies listed in Table 1. 
After typically 10 minutes of data taking, the remaining beam was 
discarded, and the cycle was repeated. 

The target and detector used for this experiment are the same as
described in Ref.\ [1], and a detailed account of the apparatus can be
found in Ref.\ [2]. The internal polarized target consisted of an 
open-ended 25\cm long storage cell of 12\mm diameter and 25\mum wall 
thickness. The cell is coated with teflon to avoid depolarization 
of atoms colliding with the wall. 
During data taking, the target polarization $\vec{Q}$ is changed every
2\s pointing in sequence, up, down ($\pm$ y), left, right ($\pm$ x), 
and along, opposite to the beam direction ($\pm$ z). The magnitude of 
the polarization is the same within $\pm$ 0.005 for all orientations [3,4]. 

The detector arrangement consists of a stack of scintillators and wire
chambers, covering a forward cone with a 30\grad opening angle. The
scintillators are capable of stopping protons from $pp \rightarrow pp\pi^0$, 
thus measuring their energy. From the time of flight and the relative energy 
deposited in the layers of the detector, the outgoing charged particles 
are identified as protons. From the direction and energy of the two 
protons in the final state, the mass $m_x$ of the undetected particle 
is calculated. For an event of interest, this mass has to equal 
the $\pi^0$ mass within the mass resolution of the experiment [1]. 
Background arises from reactions in the walls of the target cell 
[2] and, to a lesser degree, from a $\approx 1\percx$ impurity in the 
target gas. Background is rejected by a condition on the relative 
angles of the detected protons and by a cut on the transverse 
distribution of the reaction vertices. A measurement with a nitrogen 
target matches the shape of the $m_x$ distribution seen with the 
H target, except for the $\pi^0$ peak, and is used to subtract 
background which remains under the $\pi^0$ mass peak (between 5
and 10\percx). Within statistics, the background shows no spin 
dependence. The validity of the background subtraction was tested 
by varying the range of accepted masses $m_x$. 

Non-vertical beam polarization is achieved by two spin-rotating
solenoids located in non-adjacent sections of the six-sided Cooler. 
The vertical and longitudinal components of the beam polarization 
$\vec{P}$ at the target are about equal, with a small sideways 
component. Since the solenoid fields are fixed in strength, the 
exact polarization direction depends on beam energy after acceleration. 
Only either the vertical or the longitudinal component contributes to 
the total cross section, depending on the orientation of the target 
polarization. In alternating measurement cycles, the sign of the beam 
polarization is reversed. More details on the preparation of non-vertical 
beam polarization in a storage ring can be found in Refs.\ [2,4].

Data are acquired for all 12 possible polarization combinations of beam 
($+,-$) and target ($\pm$ x, $\pm$ y, $\pm$ z). The known spin correlation 
coefficients of proton-proton elastic scattering [5] are used to monitor 
beam and target polarization, concurrently with the acquisition of 
$pp \rightarrow pp\pi^0$ events. To this end, coincidences between two 
protons exiting near $\Theta_{lab} = 45\gradx$ are detected by two 
pairs of scintillators placed, behind the first wire chamber at
azimuthal angles $\pm 45\gradx$ and $\pm 135\gradx$. From this 
measurement, the products $P_yQ_y$ and $P_zQ_z$ of beam and target 
polarization are deduced. Evaluating the cross ratio of the four
$pp \rightarrow pp\pi^0$ yields with $\pm P$ and $\pm Q_y$, dividing 
by the measured $P_yQ_y$, results in \dsTx/\sigtotx. Analogously,
\dsLx/\sigtot follows from the yields with $\pm P$ and $\pm Q_z$, 
using the measured $P_zQ_z$. 

At 375 and 400\MeVx, two data runs, separated by five months, 
were combined. The detector acceptance is less 
than 100\perc because of a hole in the center of the detector stack which
accommodates the circulating beam, and excludes particles with 
$\Theta_{lab} < 5\gradx$. The resulting loss of events (about 30\perc 
at the lowest energy) is taken into account by applying a correction 
which arises from a difference in detector acceptance for Ss, 
Ps and Pp final states [1]. This correction turns out to be small 
(between  0.009 and  0.058 for \dsTx/\sigtotx, and between
0.007 and 0.030 for \dsLx/\sigtotx).

The final results are listed in Table 1. The errors include counting
statistics, and an uncertainty for background subtraction and hole
correction. The latter two were estimated by a reasonable variation 
($\pm 25\percx$) in the amount of background and in the size of 
the hole correction. The \dsT/\sigtot results from this experiment 
were consistent with those obtained in an earlier experiment [1]. 
The weighted averages of the two data sets are shown in column 7 
of Table 1. The data in the last two columns of Table 1 represent 
all experimental information on $pp \rightarrow pp\pi^0$ 
spin--dependent total cross sections and are shown in Fig.\ 1.

Most of the theoretical work on $pp \rightarrow pp\pi^0$ so far deals 
with the lowest partial wave (Ss). An exception is the meson-exchange 
model of the J\"ulich group [6] which contains the higher partial waves 
needed to address polarization observables. This calculation includes 
off-shell pion rescattering and the exchange of heavy mesons, and 
provides a good fit to the Ss part of the total cross section close 
to threshold, or for $\eta < 0.5$, where $\eta$ is the maximum 
center-of-mass pion momentum divided by the $\pi^0$ mass. 
The agreement is not as good for the 
polarization observables measured here (Fig.\ 1) which are sensitive 
to contributions from Ps and Pp partial waves.

As pointed out in the first part of this paper, the present measurement
allows a model-free statement about the Ps and Pp contributions to
$pp \rightarrow pp\pi^0$. Eq.\ 3 directly yields the relative Ps strength, 
$\sigma_{Ps}/\sigtotx$ The result is shown in Fig.\ 2. The Ps strength is 
larger than previously thought [8] (see below).

Phase space arguments together with the properties of spherical Bessel
functions for small arguments lead to the expectation that $\sigma_{Ps}$ 
is proportional to $\eta^6$. In order to test this prediction, we have 
to multiply our experimental $\sigma_{Ps}/$\sigtot by the total cross
section. At T = 325\MeV an accurate value for the total cross section 
exists (\sigtot = 7.70 $\pm$ 0.26\mubarnx, [7]). However,
at higher energies data are few and of poor quality, and a good
measurement of \sigtot between 0.3 and 1\GeV would certainly be a 
much--needed addition to the $pp \rightarrow pp\pi^0$ data base. 
For the present purpose we use a smooth approximation to the world's 
data, obtaining \sigtot = 17, 40, and 86\mubarnx, for T = 350, 375,
and 400\MeVx, respectively. The resulting values for
$\sigma_{Ps}$ are shown in Fig.\ 3. The dashed line in Fig.\ 3 
represents the fit with  
$\sigma_{Ps} = \eta^6\cdot (33.8 \pm 1.3)\mubarnx$. The error of the
scaling factor  
takes into account the errors of the data, and the uncertainty 
of \sigtot at 325\MeVx. One can also combine 
the expressions for $^3\sigma_1$ in Eqs.\ 1,2 to obtain 
an equation for ($\sigma_{Pp} - \hat{\sigma}_{Pp}$)/\sigtot in terms of  
\dsLx/\sigtotx. Multiplying the result by \sigtotx, yields the Pp partial 
cross section, $\sigma_{Pp} - \hat{\sigma}_{Pp}$, shown in Fig.\ 4. 
The dashed line in Fig.\ 4 represents a fit with  
the expected energy dependence for Pp waves, 
$(\sigma_{Pp} - \hat{\sigma}_{Pp}) = \eta^8\cdot (53.9 \pm 3.4)\mubarnx$. 
The solid lines in Figs.\ 3,4 are obtained with the J\"ulich  
meson--exchange model [6]. 
One sees that the Ps strength which, in this model, is dominated by
the role of the $\Delta$ (a nucleon excited state) is underestimated 
by about a factor of 3 (Fig.\ 3), while for ($\sigma_{Pp} - \hat{\sigma}_{Pp}$) 
the discrepancy is less but the energy dependence is not
correctly reproduced by the model (Fig.\ 4).

In the absence of polarization observables, some information on
individual partial--wave contributions can be extracted from a 
measurement of the (unpolarized) cross section as a function of 
the relative energy of the two nucleons in the final--state, 
assuming that the corresponding energy distributions of single 
partial waves are known. As our data indicate, power laws in $\eta$ 
seem adequate for the Ps and Pp partial waves, but the energy 
dependence of the Ss partial wave is dominated by the final--state
interaction, and thus requires model--dependent input. Such an 
analysis has been carried out for $pp \rightarrow pp\pi^0$ at 310\MeV 
($\eta = 0.451$) [8]. In  that work a significant  
contribution was found from an Sd-Ss 
interference term. This conclusion was reached, however, setting 
$\sigma_{Ps} = 0$, while from the present data, using an $\eta^6$ 
energy dependence, we estimate that, at 310\MeVx, $\sigma_{Ps}$ contributes 
about 6\perc to \sigtotx. 

In summary, we have measured both spin--dependent total cross sections
in $pp \rightarrow pp\pi^0$. These polarization observables allow a 
model-free determination of individual partial wave contributions. 
We show in an example that this can provide a test of individual
components of a model (Figs.\ 3,4). The availability of polarization 
observables is crucial when carrying out a partial wave study, 
since an analysis based only on the unpolarized differential cross 
section is subject to model dependence.

We thank Dr.\ Ch.\ Hanhart for supplying us with the numerical values
for the calculation shown in Figs.\ 1,3 and 4. This work has been 
supported by the US National Science Foundation under Grants PHY96-02872, 
PHY95-14566, PHY97-22556, and by the US Department of Energy under
Grant DOE-FG02-88ER40438.

\vspace{10mm}

{\parindent0mm \large \bf Table 1} \\
Bombarding energy T, maximum pion center-of-mass momentum $\eta$ in
units of the pion mass, integrated luminosity, the products 
$P_yQ_y$ and $P_zQ_z$ of beam and target polarization, and the 
measured  \dsTx/\sigtot and \dsLx/\sigtot are listed. 
Column 7 lists the values for \dsTx/\sigtot from this experiment 
(column 6) combined with the results from an earlier measurement 
with purely vertical beam polarization [1].

\begin{center}
\begin{tabular}{|c|c|c|c|c|c|c|c|} \hline
T & $\eta$ & $\int$Ldt & $P_yQ_y$ & $P_zQ_z$ & \dsTx/\sigtot & \dsTx/\sigtot &
\dsLx/\sigtot \\ 
\MeV & & $(\nbarnx)^{-1}$ & & & this exp & incl.\ [1] & this exp\\ \hline
325.6 & 0.560 & 3.0 & 0.333(2) & 0.296(3) & $-1.155 \pm 0.106$ &
$-1.078 \pm 0.063$ & $1.623 \pm 0.116$ \\ \hline
350.5 & 0.707 & 1.4 & 0.316(3) & 0.267(5) & $-0.524 \pm 0.099$ &
$-0.484 \pm 0.067$ & $1.277 \pm 0.119$ \\ \hline
375.0 & 0.832 & 4.1 & 0.333(2) & 0.266(4) & $-0.239 \pm 0.039$ &
$-0.274 \pm 0.021$ & $0.676 \pm 0.049$ \\ \hline
400.0 & 0.948 & 1.1 & 0.289(4) & 0.203(8) & $-0.088 \pm 0.065$ &
$-0.076 \pm 0.038$ & $0.590 \pm 0.093$ \\ \hline
\end{tabular}
\end{center}

\section*{Figure Captions}
\begin{enumerate}
\item[Fig.\ 1.]  The $pp \rightarrow pp\pi^0$ spin--dependent total cross 
      sections \dsTx/\sigtot (column 7 in Table 1) and \dsLx/\sigtot 
      (column 8) as a function of $\eta$, the maximum pion 
      center-of-mass momentum 
      in units of the pion mass. The curves represent the predictions 
      by the J\"ulich meson--exchange model [6].
\item[Fig.\ 2.] Measured, relative Ps strength, $\sigma_{Ps}$/\sigtot, of 
      the reaction $pp \rightarrow pp\pi^0$ as a function of $\eta$.
\item[Fig.\ 3.] The total cross section $\sigma_{Ps}$ for $pp
      \rightarrow pp\pi^0$ as a function of $\eta$. The dashed line 
      represents the best fit with  $\sigma_{Ps}$ proportional to  $\eta^6$.
      The solid line is the Ps total cross section calculated from the
      J\"ulich  meson-exchange model [6]. 
\item[Fig.\ 4.] The total cross section $\sigma_{Ps} - \hat{\sigma}_{Pp}$ for
      $pp \rightarrow pp\pi^0$ as a function of $\eta$. The dashed line 
      represents the best fit with $\sigma_{Ps} - \hat{\sigma}_{Pp}$
      proportional to $\eta^8$. The solid line is the prediction of the 
      J\"ulich meson--exchange model [6].
\end{enumerate}


\newpage

\begin{thebibliography}{1}
\bibitem[1]{Mey98} 
        H.O.\ Meyer et al.\ Phys.\ Rev.\ Lett.\ {\bf 81}:3096, 1998.
\bibitem[2]{Rin98} 
        T.\ Rinckel et al.\ Submitted to Nul.\ Instr.\ Meth. 
\bibitem[3]{Rat98} 
        F.\ Rathmann et al.\ Phys.\ Rev.\ {\bf C58}:658, 1998.
\bibitem[4]{Lor98} 
        B.\ Lorentz et al.\ Submitted to Phys.\ Rev.
\bibitem[5]{BvP98} 
        B.\ v.\ Przewoski et al.\ Phys.\ Rev.\ {\bf C58}:1897, 1998.
\bibitem[6]{Han98} 
        C.\ Hanhart, J.\ Haidenbauer, O.\ Krehl and J.\ Speth.\ 
        Phys.\ Lett.\ {\bf B444}:25, 1998.
\bibitem[7]{Mey92} 
        H.O.\ Meyer et al.\ Nucl.\ Phys.\ {\bf A539}:633, 1992.
\bibitem[8]{Zlo98} 
        J.\ Z{\l}oma\'{n}czuk et al.\ Phys.\ Lett.\ {\bf B436}:251, 1998.
\end{thebibliography}
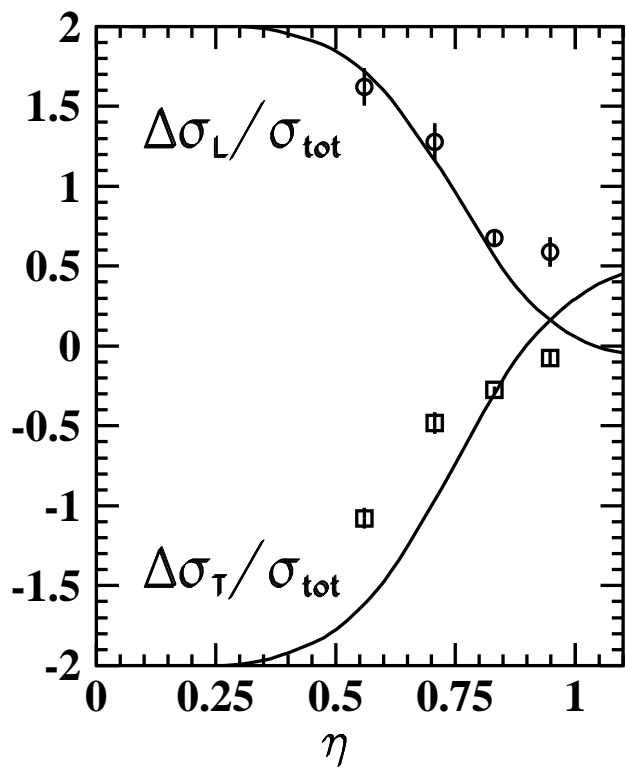
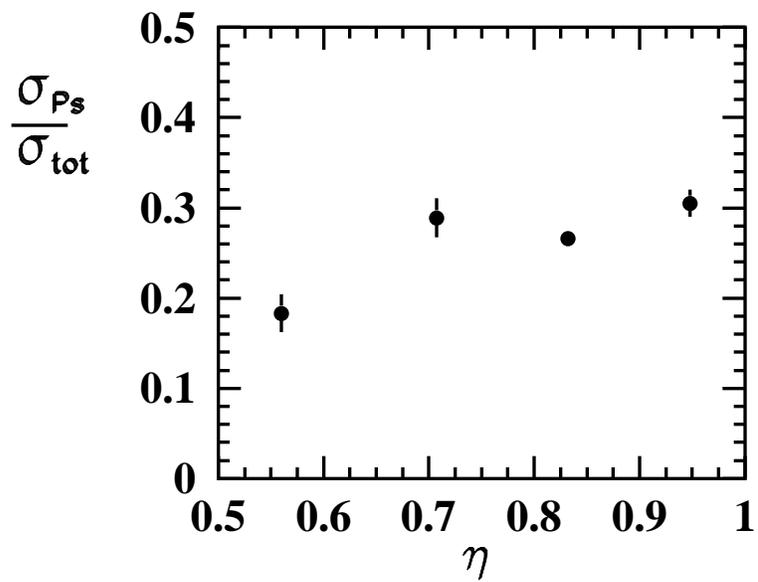
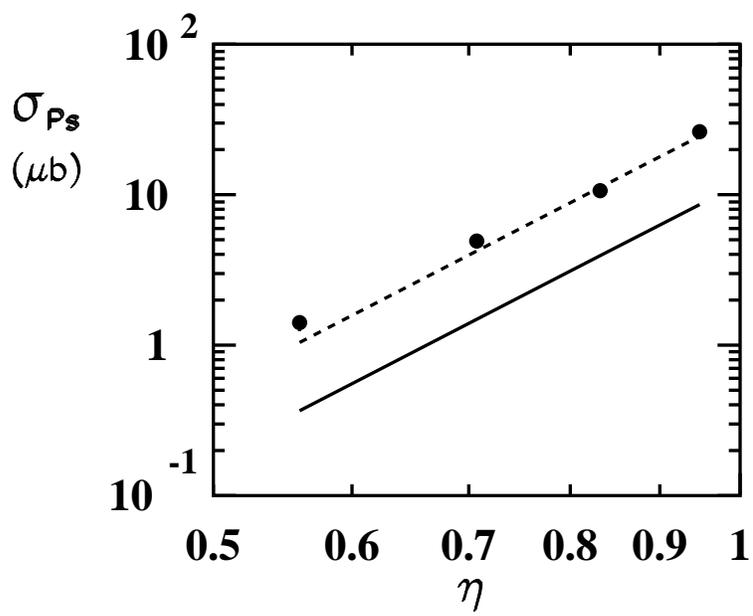
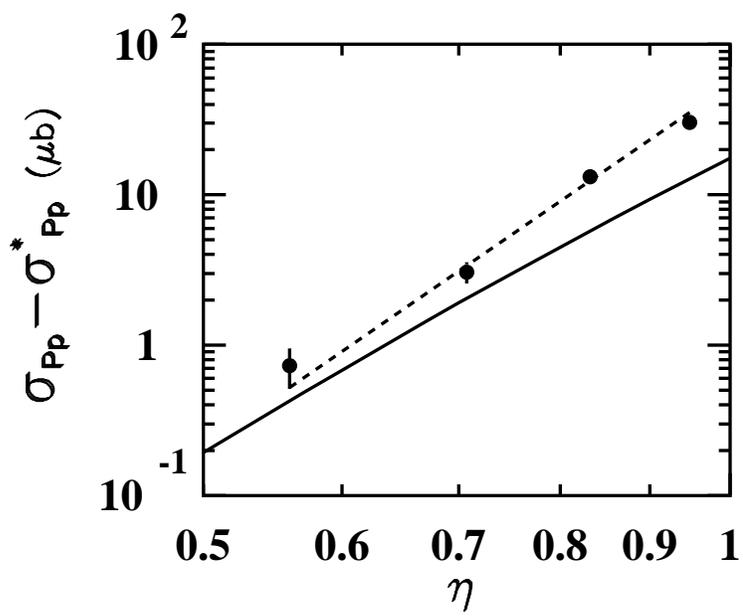
\end{document}